\documentclass[aps,prl,twocolumn,superscriptaddress,showpacs]{revtex4}

\usepackage{qsymbols}
\usepackage[dvips]{graphicx}
\usepackage{eqnarray}
\usepackage[colorlinks=true,urlcolor=blue,citecolor=blue,linkcolor=black,breaklinks=true,dvips]{hyperref} % pour mettre les liens hypertexte

\begin{document}

% Use the \preprint command to place your local institutional report
% number in the upper righthand corner of the title page in preprint mode.
% Multiple \preprint commands are allowed.
% Use the 'preprintnumbers' class option to override journal defaults
% to display numbers if necessary
%\preprint{}

%Title of paper
\title{Local contact stress measurements at a rough interface}
%\title{Contact stress field measurement at a rough interface}

% repeat the \author .. \affiliation  etc. as needed
% \email, \thanks, \homepage, \altaffiliation all apply to the current
% author. Explanatory text should go in the []'s, actual e-mail
% address or url should go in the {}'s for \email and \homepage.
% Please use the appropriate macro foreach each type of information

% \affiliation command applies to all authors since the last
% \affiliation command. The \affiliation command should follow the
% other information
% \affiliation can be followed by \email, \homepage, \thanks as well.
\author{J. Scheibert}
%\email[]{Your e-mail address}
%\homepage[]{Your web page}
%\thanks{}
%\altaffiliation{}
\affiliation{Laboratoire de Physique Statistique de l'ENS, UMR 8550 CNRS/ENS/Universit\'e Paris 6/Universit\'e Paris 7, 24 rue Lhomond, 75231 Paris, France}
\author{A. Prevost}
%\email[]{Your e-mail address}
%\homepage[]{Your web page}
%\thanks{}
%\altaffiliation{}
\affiliation{Laboratoire de Physique Statistique de l'ENS, UMR 8550 CNRS/ENS/Universit\'e Paris 6/Universit\'e Paris 7, 24 rue Lhomond, 75231 Paris, France}
\author{J. Frelat}
%\email[]{Your e-mail address}
%\homepage[]{Your web page}
%\thanks{}
%\altaffiliation{}
\affiliation{Laboratoire de Mod\'elisation en M\'ecanique, UMR 7607 CNRS/Universit\'e Paris 6, 4 place Jussieu, 75252 Paris, France}
\author{P. Rey}
%\email[]{Your e-mail address}
%\homepage[]{Your web page}
%\thanks{}
%\altaffiliation{}
\affiliation{CEA-LETI, 17 rue de Martyrs, F38054 Grenoble Cedex 09, France}
\author{G. Debr\'egeas}
%\email[]{Your e-mail address}
%\homepage[]{Your web page}
%\thanks{}
%\altaffiliation{}
\affiliation{Laboratoire de Physique Statistique de l'ENS, UMR 8550 CNRS/ENS/Universit\'e Paris 6/Universit\'e Paris 7, 24 rue Lhomond, 75231 Paris, France}

%Collaboration name if desired (requires use of superscriptaddress
%option in \documentclass). \noaffiliation is required (may also be
%used with the \author command).
%\collaboration can be followed by \email, \homepage, \thanks as well.
%\collaboration{}
%\noaffiliation

\date{\today}

\begin{abstract}
An original MEMS-based force sensing device is designed which allows to measure spatially resolved normal and tangential stress fields at the base of an elastomeric film. This device is used for the study of the contact stress between a rough film and a smooth glass sphere under normal load. The measured profiles are compared to Finite Elements Method calculations for a smooth contact with boundary conditions obeying Amontons-Coulomb's friction law. The accuracy of the measurements allows to discriminate between dry and lubricated contact conditions and to evidence load-dependent deviations from Amontons-Coulomb's profiles. These deviations are qualitatively interpreted by taking into account the finite compliance of the micro-contact population. 
\end{abstract}

% insert suggested PACS numbers in braces on next line
% insert suggested keywords - APS authors don't need to do this
%\keywords{}

%\maketitle must follow title, authors, abstract, \pacs, and \keywords
\maketitle

% body of paper here - Use proper section commands

Knowledge of the stress field at the contact between two solids is of considerable interest in numerous contexts such as mechanical engineering, seismology or solid friction. Starting with the classical analysis of Hertz this problem has been extensively studied theoretically in various geometries for simple boundary conditions \cite{Hertz-1896, Spence-JElasticity-1975, Johnson-CUP-1985}. These calculations generally assume smooth interfaces which makes them unapplicable to most practical situations: real solids are generally rough at the micrometer scale. The apparent contact is therefore made of a large number of isolated load-bearing micro-contacts, which is expected to modify the overall stress distribution as predicted in \cite{Greenwood-Tripp-JAppMech-1967, Persson-PRL-2001}. Exhibiting such deviations requires to access the stress field with a spatial resolution below the contact size. In recent years, Micro Electro Mechanical Systems (MEMS) have emerged as convenient tools for such stress measurements \cite{Kane-Cutkosky-Kovacs-SensorsActuatorsA-1996, Leineweber-Pelz-Schmidt-Kappert-Zimmer-SensorsActuatorsA-2000}. Their quasi-monocristalline Silicon structure offers a high linearity and a low hysteresis response. Their production processes, borrowed from micro-electronic technology, also allow for progressive miniaturization and production of arrays of sensors. 

\begin{figure}
\includegraphics[width=\columnwidth]{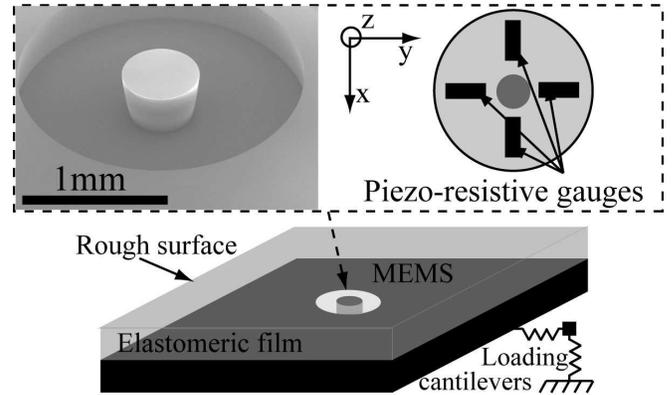}
\caption{Schematics of the stress sensor. The MEMS sensitive part is a rigid cylinder (diameter $550 \, \mu m$, length $475 \, \mu m$) attached to a suspended circular Silicon membrane (radius $1 \, mm$, thickness $100 \, \mu m$) whose deformation is measured via four couples of piezo-resistive gauges. It enables to measure the applied forces along three orthogonal directions at the base of a nominally flat rough PDMS film (thickness $2 \, mm$, lateral dimensions $50 \times 50 \, mm$). The macroscopic normal and tangential loads applied on the film are measured through the extension of two orthogonal loading cantilevers (normal stiffness $641 \pm 5 \, N.m^{-1}$, tangential stiffness $51100 \pm 700 \, N.m^{-1}$) by capacitive position sensors (respectively MCC30 and MCC5, Fogale nanotech).\label{schemamanip}}
\end{figure}

In this Letter, we report on a new stress sensing device based on a triaxial MEMS force-sensor embedded at the base of rough, nominally flat elastomeric film (Fig. \ref{schemamanip}). The normal and tangential stress profiles are measured in the case of a sphere-on-plane contact under normal load. The high accuracy of the measurements allows for a direct comparison to elastic calculations under various mechanical boundary conditions. 

The elastomeric material is a cross-linked poly\-di\-methyl\-siloxane (PDMS) (Sylgard 184, Dow Corning) of Young's modulus $2.2 \pm 0.1 \, MPa$, and of Poisson ratio $0.5$ \cite{PolyDataHandbook-OUP-1999}. No measurable stress relaxation being observed after a sudden loading, the PDMS can be considered as purely elastic. The film is obtained by pouring the cross-linker/PDMS melt on the MEMS into a parallelepipedic mold covered with a Poly\-Methyl\-Meth\-Acrylate plate roughened by abrasion with an aqueous solution of Silicon Carbide powder (mean diameter $37 \, \mu m$). After curing and demolding the resulting $rms$ surface roughness is measured with an interferential optical profilometer (M3D, Fogale Nanotech) to be $\rho=1.82 \pm 0.10 \, \mu m$. This roughness is sufficient to avoid any measurable pull-off force against smooth glass substrates, as discussed in \cite{Fuller-Tabor-ProcRSocLondA-1975}. 

\begin{figure}
\includegraphics[width=\columnwidth]{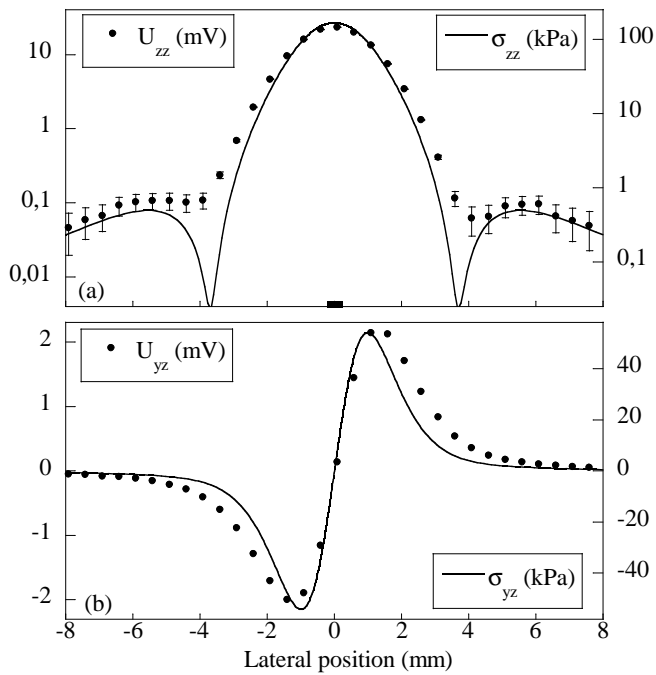}
\caption{Radial stress profiles under normal loading by a rigid thin rod. The measured tension $U$ ($`(!)$) is compared to FEM results (solid line) for a normal load of $1 \, N$ for (a) the normal and (b) the tangential stress. The error bars, when visible, represent the electronic noise. The black rectangular patch on (a) represents the rod diameter.\label{IP-PRL1}}
\end{figure}

The force sensing device is first calibrated by indentating the film surface with a rigid rod of diameter $500 \, \mu m$ (details of the complete calibration procedure are given in \cite{Scheibert-Prevost-Frelat-Rey-Debregeas-tobepublished}). In this contact geometry the sensor measurement is found to be linear with the applied load. By successively varying the position of the contact with regards to the sensor symmetry axis along the $y$ direction, one can reconstruct both the normal ($zz$ subscript) and tangential ($yz$ subscript) output voltage radial profiles $U$ (Fig.\ref{IP-PRL1}). They are then compared to the results of Finite Elements Method (FEM, Software Castem 2007) calculations for the stress $\sigma$ at the base of an axi-symmetrical elastic film (of same elastic moduli) perfectly adhering to its base and submitted to a prescribed normal displacement over a central circular area \footnote{Such results could have been obtained semi-analytically for frictionless conditions by using the model developed in \cite{Fretigny-Chateauminois-JPhysD-2007} but FEM calculations have been preferred here because they allow for variable boundary conditions.}. As expected for contacts of dimensions smaller than the film thickness, the stress calculated at the base of the film is found to be insensitive to the friction boundary conditions. For each stress component the relationship between the stress field $\sigma$ and the output tension $U$ is assumed to be
\begin{equation}
U(x,y)=A_0 . G \otimes \sigma (x,y) \label{convol}
\end{equation}
where $A_0$ is a conversion constant (expressed in $mV / Pa$), $G$ is a normalized apparatus function and $\otimes$ is a convolution product. 

$G_{zz}$ and $G_{yz}$ are obtained by numerical integration of Eq.\ref{convol} in Fourier space. The resulting functions exhibit a bell shape with typical width of the order of $600 \, \mu m$ comparable to the MEMS lateral dimension. $G_{zz}$ has proven to be adequate for all contact situations tested, therefore validating the hypothesis underlying Eq.\ref{convol}. In contrast the tangential output of the MEMS appears to be sensitive to the presence of normal stress gradients over the MEMS size, which prevents a solid determination of $G_{yz}$ and hence robust measurements of the tangential stress \footnote{This limitation is expected to vanish upon miniaturization.}. Still, comparative information can be extracted when considering situations exhibiting low normal stress gradients, such as contacts against spheres having a large radius of curvature as studied in this Letter. For this subclass of situations an effective apparatus function having the same spatial distribution as $G_{yz}$ but an amplitude increased by $20\%$ is used, for it provides satisfying agreement between experimental and calculated tangential stress profiles over the whole range of applied loads. 

Contact is made with an optical plano-convex spherical glass lens (radius of curvature $128.8 \, mm$). Both the glass and the PDMS surfaces are passivated using a vapor-phase silanization procedure which reduces and homogenizes the surface energy. The contacts are obtained using the following loading sequence. The substrate is pressed against the film up to the prescribed load $P$ within $2 \, \%$ relative error. Due to the associated tangential displacement of the extremity of the normal cantilever, a significant tangential load $Q$ is induced. From this position, the contact is renewed by manual separation which results in a much smaller but finite $Q$. The substrate is eventually translated a few $\mu m$ tangentially down to $Q=0$. Both the surface treatment and the loading sequence yield an excellent measurement reproducibility. As for the rod indentation the radial profiles are derived assuming a homogeneity of the surface properties of the film from a series of $33$ contacts whose centers lie every $0.5 \, mm$ along the $y$ direction. These profiles divided by $A_0$ are denoted as $S$ and have the dimension of a stress.

For quantitative comparisons, FEM calculations are carried out for a frictional sphere-on-plane contact with the same geometry. Both contacting surfaces are taken to be smooth. The interface is assumed to obey a local Amontons-Coulomb's law of friction which states that no slip occurs wherever the ratio of tangential over normal stress remains smaller than the static friction coefficient $\mu$. The normal displacement of the rigid elastic sphere is prescribed. Both solids are discretized with a uniform mesh size of $50 \, \mu m$ and the normal displacement of the rigid elastic sphere is prescribed. The contact conditions are satisfied using a double Lagrange multiplier implying that both surfaces are slave and master. The normal load is reached step-by-step: at each step an iterative Newton-Raphson method is used to satisfy both the unilateral contact and the friction law.

\begin{figure}
\includegraphics[width=\columnwidth]{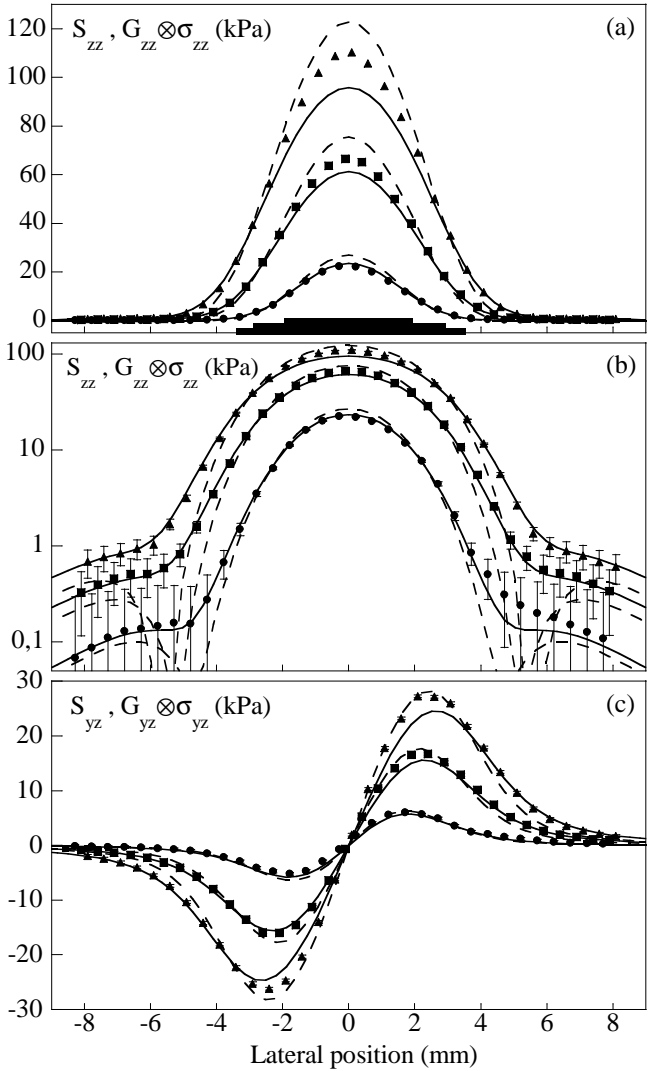}
\caption{Measured stress profiles ($S$) under normal loading by a rigid sphere ($P=0.34 \, N$ ($\bullet$), $1.37 \, N$ ($`[!]$), $2.75 \, N$ ($\blacktriangle$)). Comparison is made with $G \otimes \sigma$ for $\mu=0$ (solid line) and $\mu=\infty$ (dashed line). (a), (b) Normal stress in linear and semi-logarithmic scale respectively. (c) Tangential stress. The black rectangular patches on (a) represent the contact diameters obtained from the FEM calculations for $\mu=0$. \label{FIG-PRL1-2}}
\end{figure}

Figure \ref{FIG-PRL1-2} compares the $S$ and $G \otimes \sigma$ profiles for three values of $P$. Within the error bar the measured profiles are bounded by both numerical profiles obtained with $\mu=0$ and $\mu=\infty$ over the whole spatial range and over 3 orders of magnitude, as clearly displayed on Fig.\ref{FIG-PRL1-2}b. In the outer region of the contact $S$ is systematically very close to the frictionless profile whereas at the center it increasingly departs from it with the load, as discussed further. The slight asymmetry of the tangential stress profiles (Fig.\ref{FIG-PRL1-2}c) can be related to some irreversible micro-slip which occurs during the loading sequence.

\begin{figure}
\includegraphics[width=\columnwidth]{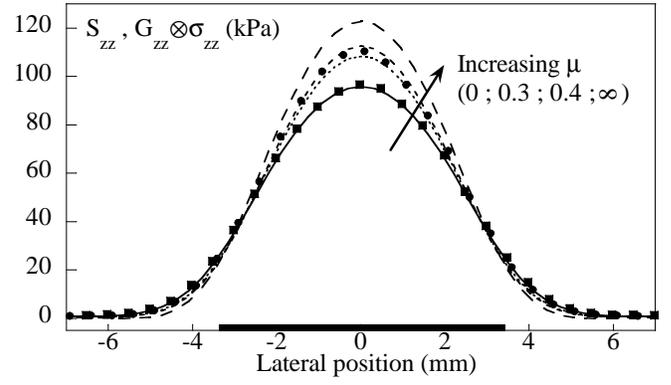}
\caption{Measured normal stress profile ($S$) under normal loading ($P=2.75 \, N$) by a rigid sphere for both dry ($`(!)$) glycerol lubricated ($`[!]$) contacts. Shown in solid and dashed lines are the $G \otimes \sigma$ profiles for 4 values of $\mu$ ($0$, $0.3$, $0.4$, $\infty$). The black rectangular patch represent the contact diameter obtained from the FEM calculations for $\mu=0$. \label{P2-PRL1-toutesf}}
\end{figure}

The effect of friction can be probed by performing similar measurements when the contact is lubricated. A glycerol droplet is inserted at the interface prior to loading, allowing for a complete relaxation of the interfacial tangential stress. This is done for two limiting loads ($P=0.69 \, N$ and $P=2.75 \, N$). The lubricated profiles are characterized by both a larger spatial extent and a lower maximum amplitude than what is found for the dry ones. In addition the lubricated profiles are found to follow almost perfectly the frictionless $G \otimes \sigma$ curves (as shown for $S_{zz}$ on Fig.\ref{P2-PRL1-toutesf} at $P=2.75 \, N$).

\begin{figure}
\includegraphics[width=\columnwidth]{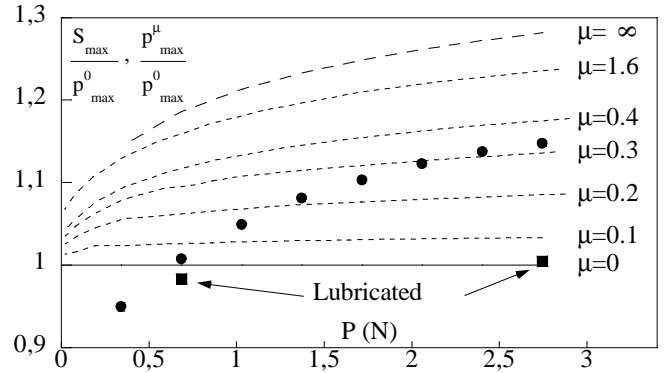}
\caption{Load dependence of the maximum normal stress $S_{max}$ at the base of the elastic film as a function of $P$, for dry ($`(!)$) and lubricated ($`[!]$) contacts), together with the maximum calculated normal stress $p^{\mu}_{max}$ for various values of $\mu$. All values are normalized by $p^0_{max}$. \label{Pmax-PRL1}}
\end{figure}

For dry contacts and sufficiently high loads, we find similarly that the normal stress profiles can be strictly bounded by the calculated ones for two very close values of $\mu$, hence providing an interval for an effective friction coefficient $\mu_e$ (for instance for $P=2.75 \, N$, $\mu_e$ is in the $[0.3,0.4]$ range as shown on Fig.\ref{P2-PRL1-toutesf}) which value is found to increase with $P$. However, for small enough loads the experimental profiles are beyond the possible range for smooth contacts and $\mu_e$ cannot be defined. This behavior can be well caught by plotting for each load the maximum value of the pressure profiles $S_{max}$ normalized by $p^0_{max}=G_{zz} \otimes \sigma(0)$ expected for a smooth frictionless contact (Fig.\ref{Pmax-PRL1}). It is compared to $p^{\mu}_{max} / p^0_{max}$ calculated for $\mu$ varying from $0$ to $\infty$. Whenever possible the value of $\mu_e$ can be directly read from Fig.\ref{Pmax-PRL1}. Clearly the experimental data cannot be described within the Amontons-Coulomb's framework using a single friction coefficient. 

The values of $\mu_e$ can be compared to the actual static friction coefficient $\mu_s$ defined as the ratio of the tangential over the normal loads at the onset of macroscopic sliding. $\mu_s$ has been measured for a driving velocity $v=100\mu m/s$ of the rigid base of the sensor and is found to decrease from $1.8$ to $1.5$ as the load is increased from $0.34 \, N$ to $2.75 \, N$, a behavior which is usually attributed to the finite adhesion energy of the interface \cite{Carbone-Mangialardi-JMechPhysSolids-2004}. We observe an opposite load-dependence for $\mu_e$, whose understanding implies therefore to reconsider the Amontons-Coulomb's assumption of a smooth undeformable interface.

A description of the normal compressibility of a rough interface \cite{Brown-Scholz-JGR-1985, Benz-Rosenberg-Kramer-Israelachvili-JPhysChemB-2006, Persson-PRL-2007} has first been proposed by Greenwood and Williamson \cite{Greenwood-Williamson-ProcRSocLondA-1966} by modeling the interface as an ensemble of spherical asperities of equal radius whose heights are statistically distributed. They showed that the true area of contact is proportional to the confining load thus providing an understanding of the quasi-independence of the static friction coefficient with the apparent area of contact. When adapted to a sphere-on-plane contact \cite{Greenwood-Tripp-JAppMech-1967} this model predicts an increase of the apparent radius of contact with respect to Hertz's theory and a decrease in the maximum normal stress. These effects are expected to be enhanced for small loads as the normal displacement of the rigid substrate becomes comparable to the thickness $\rho$ of the rough layer. This could explain why $S_{max} / p^0_{max}$ falls below $1$ for vanishingly small applied loads.

At high enough load, the fact that the lubricated profiles are very close to those calculated for a smooth frictionless contact suggests that these normal compressibility effects are negligible. The finite shear stiffness of the interface, evidenced experimentally in \cite{Berthoud-Baumberger-ProcRSocA-1998, Bureau-Caroli-Baumberger-ProcRSocLondA-2003}, is then expected to be primarily responsible for the drop from $\mu_s$ to $\mu_e$. It allows for a partial relaxation of the tangential stress before the onset of slippage, which qualitatively translates into an apparent reduction of the friction coefficient. This reduction should be larger the smaller the load, when the relaxed tangential displacement becomes comparable to the slip amplitude expected for a frictionless smooth contact. This behavior is consistent with what is observed on Fig.\ref{Pmax-PRL1}. At this point a quantitative description of the mechanical state of the system should involve an elasto-plastic like friction law accounting for the ability of the asperities to deform reversibly before slippage.

The stress sensor described in this Letter has proven to be well-suited for the study of stress fields in centimeter-sized contacts. The measurements are accurate enough to discriminate between lubricated and dry contacts and to evidence deviations from Amontons-Coulomb's model of friction induced by a micron-sized interfacial roughness. These measurements can then provide a fine test for any mechanical model for the frictional interface. Many other aspects of friction and contact mechanics, such as the dynamical frictional regimes or the history-dependence of a contact submitted to an oscillatory tangential load below the sliding threshold, are likely to be probed with the same device by applying controlled tangential load sequences. This device might also be of interest in domains such as rheology or adhesion where accurate spatially resolved stress measurements at interfaces are needed.

\begin{acknowledgments}
The authors thank A. Ponton and B. Ladoux for their help in the measurement of the PDMS Young's modulus and A. Chateauminois and C. Fretigny for fruitful discussions.
\end{acknowledgments}

% Create the reference section using BibTeX:
%\bibliography{BiblioPRL1/biblioPRL1}
  
\end{document}